\documentclass[aps,prl,twocolumn,showpacs,nofootinbib,superscriptaddress]{revtex4-2}

\usepackage{amsmath}
\usepackage{amsfonts}
\usepackage{amssymb}
\usepackage{mathtools}
\usepackage{bm}          
\usepackage{dcolumn}     
\usepackage{slashed}     
\usepackage{upgreek}     
\usepackage{bigints}     
\usepackage{graphicx}
\usepackage{epsfig}
\usepackage[dvipsnames,table]{xcolor}
\usepackage{colortbl}
\usepackage{adjustbox}
\usepackage[colorlinks=true,linkcolor=blue]{hyperref}%
\usepackage{float}
\usepackage[caption=false]{subfig}

\usepackage{array}
\usepackage{multirow}
\usepackage{makecell}
\usepackage{hhline}
\usepackage{diagbox}     

\usepackage{placeins}
\usepackage{textcomp}
\usepackage{blindtext}
\usepackage{url}
\usepackage[utf8]{inputenc}
\usepackage{docs}
\usepackage[toc,page]{appendix}
\usepackage[normalem]{ulem}
\usepackage{lipsum}
\usepackage{chngcntr}
\usepackage[shortlabels]{enumitem}
\usepackage{empheq}

\usepackage[colorlinks=true,linkcolor=blue]{hyperref}
\usepackage{cleveref}
\usepackage[toc,page]{appendix}
\usepackage{tikz}
\usepackage{tikz-feynman}
\usepackage{pifont}
\usetikzlibrary{snakes, arrows}
\tikzfeynmanset{compat=1.1.0}
\tikzset{vertex/.style={circle,draw, minimum size=1.5em}, edge/.style={->,> = latex'}}

\usepackage{silence}
\tikzfeynmanset{warn luatex=false}

\crefname{figure}{fig\,.}{figs\,.} 
\crefname{equation}{eq\,.}{eqs\,.} 
\newcommand{\stkout}[1]{\ifmmode\text{\sout{\ensuremath{#1}}}\else\sout{#1}\fi}
\definecolor{calpolypomonagreen}{rgb}{0.12, 0.3, 0.17}
\newcommand{\bea}{\begin{eqnarray}}
	\newcommand{\eea}{\end{eqnarray}}

\usepackage{xcolor}

\def\Dst{{D^*}}

\def\Dst{{D^*}}

\def\thD{{\theta_D}}
\def\thl{{\theta_\ell}}
\def\nn{\nonumber}

\begin{document}

\title{ Probing Invisible Fermions in $B \to D^{*}\ell X_{\text{inv}}$ via Angular Observables}

\author{Lipika Kolay}
\affiliation{Department of Physics, Indian Institute of Technology Guwahati, North Guwahati, Assam-781039, India}
\affiliation{Department of Physics, Indian Institute of Technology Gandhinagar, India}
\author{Soumitra Nandi}
\affiliation{Department of Physics, Indian Institute of Technology Guwahati, North Guwahati, Assam-781039, India}

\author{Shantanu Sahoo}
\affiliation{School of Physics, University of Hyderabad, Hyderabad, Telangana-500046, India}


\begin{abstract}
Semileptonic decays $B \to D^{*} \ell X_{\text{inv}}$ provide a sensitive probe of light invisible particles, such as sterile neutrinos or dark-sector fermions. Within a general weak effective theory framework, we show that a massive invisible fermion induces distinctive modifications in the angular distributions. We identify observables with enhanced sensitivity to the invisible particle mass, allowing a clear discrimination of such scenarios, and highlight angular structures that differentiate left- and right-handed lepton-dark-sector currents.
\end{abstract}

\maketitle




\paragraph{\underline{Introduction}:}

Semileptonic decays $ B \to D^{*} \ell X_{\rm inv} $ play a central role in flavour physics, providing a primary avenue for the extraction of the CKM matrix element $|V_{cb}|$ and stringent tests of the Standard Model (SM). Recent measurements from Belle and Belle-II \cite{Belle:2023bwv,Belle-II:2023okj} have significantly improved our understanding of the kinematics and angular structure of these modes. Owing to their clean theoretical interpretation and experimental accessibility, these decays offer a powerful laboratory to probe potential deviations from the SM through precision observables.

A key feature of these processes is the presence of an invisible particle in the final state, conventionally assumed to be a massless SM neutrino. Experimentally, this appears as missing energy, making the reconstruction of decay kinematics challenging. At the same time, it opens an opportunity to test the nature of the invisible sector. In particular, recent analyses of the missing mass squared distribution \cite{Belle:2023bwv} show a peak near zero, with a small but non-vanishing spread, motivating the exploration of scenarios where the missing particle may not be strictly massless.

Such an interpretation naturally arises in a wide class of beyond-the-Standard-Model (BSM) scenarios, where the invisible particle could be a light sterile neutrino or a dark-sector fermion. If present, a massive invisible particle could modify the decay kinematics and induce characteristic distortions in the angular distributions. These effects provide a complementary and largely unexplored handle to probe the nature of the invisible sector in semileptonic $B$ decays.

In this work, we investigate the impact of a light invisible fermion $\chi$ in $ B \to D^{*} \ell \chi $ decays within a model-independent effective field theory framework. Allowing for both left- and right-handed interactions, we focus on the role of the invisible particle mass in shaping angular observables and phase-space distributions. We demonstrate that specific observables exhibit enhanced sensitivity to $m_\chi$ and can discriminate between different chiral structures, thereby providing new probes of light dark-sector scenarios in flavour physics.

 
\paragraph{\underline{Theory Framework}:}
We describe the decay $ B \to D^{*} \ell X_{\rm inv} $ within a general weak effective theory framework, where $X_{\rm inv}$ denotes either a SM neutrino or a light BSM fermion $\chi$. The effective Hamiltonian for $b \to c \ell X_{\rm inv}$ (with $\ell = e, \, \mu$) transitions is
\begin{equation}\label{eq:general_Hamiltonian_b2c}
	\mathcal{H}_{\rm eff} = \frac{4G_F}{\sqrt{2}} V_{cb} \sum_i C_i^\ell \, \mathcal{O}_i^\ell \, ~,
\end{equation}
where we retain the relevant vector, scalar, and tensor operators,
\begin{align}
\label{eq:operators}
 & \mathcal{O}_{V_{1L}}^{\ell,SM} = \left(\bar{c}\gamma_{\mu} P_{L} b \right)\left( \bar{\ell} \gamma^{\mu} P_{L}\nu_{\ell} \right) \,,   \nonumber \\
 & \mathcal{O}_{V_{1L(R)}}^{\ell} = \left(\bar{c}\gamma_{\mu} P_{L} b \right)\left( \bar{\ell} \gamma^{\mu} P_{L(R)} \chi \right)\,,  \nonumber \\&
 \mathcal{O}_{V_{2L(R)}}^{\ell}  = \left(\bar{c}\gamma_{\mu} P_{R} b \right)\left( \bar{\ell} \gamma^{\mu} P_{L(R)} \chi \right)\,,  \nn \\ 
 &\mathcal{O}_{S_{1L(R)}}^{\ell} = \left(\bar{c} P_{R}  b \right)\left( \bar{\ell} P_{L(R)} \chi \right)\,, \qquad \,
 \nonumber \\&
 \mathcal{O}_{S_{2L(R)}}^{\ell} = \left(\bar{c}P_{L} b \right)\left( \bar{\ell} P_{L(R)} \chi \right) \,,  \nonumber \\ 
&\mathcal{O}_{TL(R)}^{\ell} = \left(\bar{c}\sigma_{\mu\nu} P_{L} (P_{R})  b \right)\left( \bar{\ell} \sigma^{\mu \nu } P_{L} (P_{R}) \chi \right)\,. 
\end{align}
The SM contribution corresponds to $\mathcal{O}_{V_{1L}}^{\ell,SM}$ with $C_{V_{1L}}^{\ell,SM}=1$, involving left-handed neutrinos. On the other hand, the dark fermion $\chi$ can have both chiralities, leading to additional operator structures. In this work, we consider real Wilson coefficients (WCs) and focus on the light leptons, i.e., $\ell=(e,\mu)$. For $\chi$ to appear as missing energy, its mass must satisfy
$m_\chi \leq (m_B - m_{D^{*}} - m_\ell)$. Such operators can arise in dark-sector effective theories and simplified models (see, e.g.,~\cite{Aebischer:2022wnl}). For example, the operators in eq.~\eqref{eq:general_Hamiltonian_b2c} can be matched onto the dimension-6 dark low-energy effective theory (DLEFT) basis \cite{Aebischer:2022wnl}, which extends the general SM low-energy basis
by dark-sector states and is valid below the electroweak scale with $SU(3)_c \otimes U(1)_{\rm em}$ invariance.
Such effective interactions can also arise in ultraviolet-complete frameworks. In particular, multicomponent dark matter models with scalar or vector leptoquark mediators \cite{Belfatto:2021ats}, as well as lepton-portal dark matter scenarios \cite{Okawa:2020jea, Iguro:2022tmr, Higuchi:2023kbt}, right-handed sterile neutrino \cite{Drewes:2016upu, Boyarsky:2018tvu} can generate the operators in eq.~\eqref{eq:general_Hamiltonian_b2c} at tree level with WCs of ${\cal O}(1)$ \cite{Kolay:2026mgv}. This provides a well-motivated top-down origin for the effective Hamiltonian considered in this work.

\begin{figure}[t]
	\centering
	\includegraphics[width=0.7\linewidth]{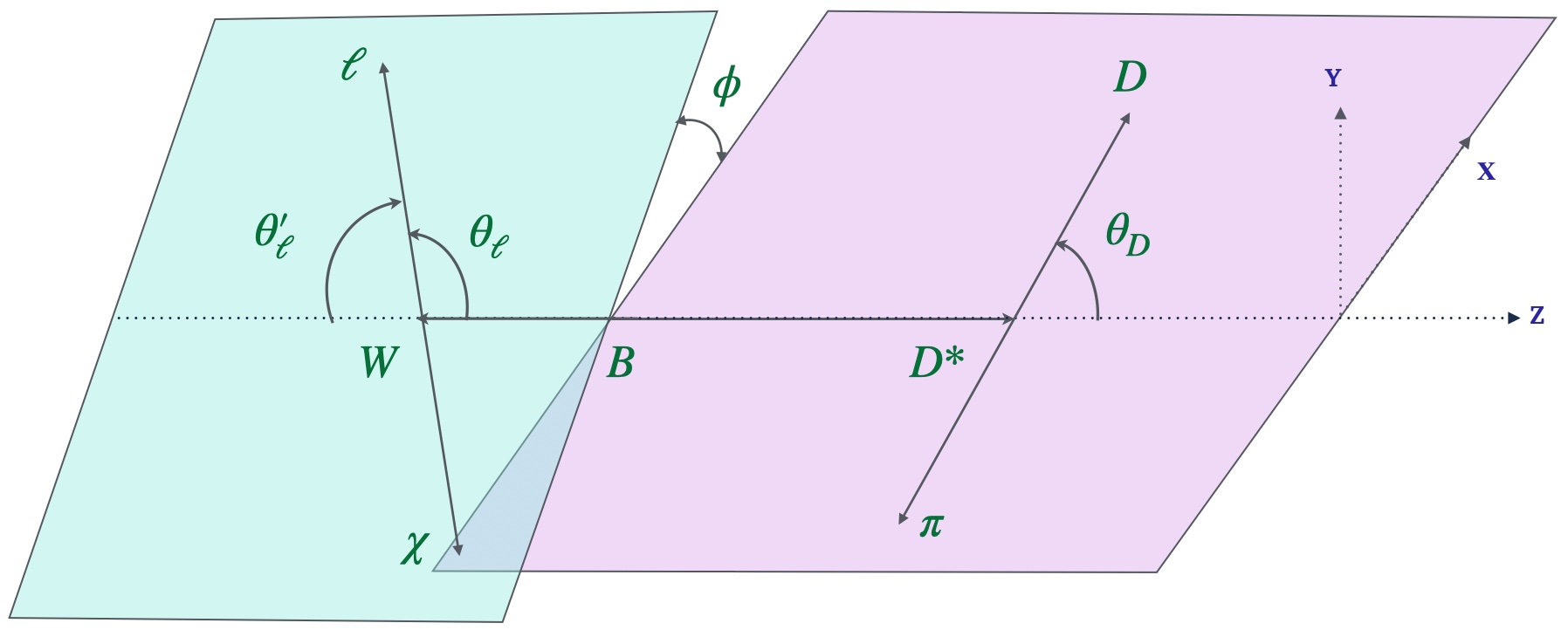}
	\caption{Schematic diagram of the decay plane of $B \to D^{*} (\to D \, \pi) \ell \chi $. }
	\label{fig:decay_plane}
\end{figure}

\paragraph{\underline{Angular Observables} :}
We study the angular observables associated with four-fold differential decay distributions of $B \to D^{*} (\to D \pi) \ell X_{\rm inv}$, where $X_{\rm inv}$ can be either a SM neutrino or a light dark-sector fermion $\chi$ with mass $ m_{\chi} $.
Since both escape detection, they appear as missing energy; however, a non-zero $m_\chi$ modifies the available phase space. In the SM, the dilepton invariant mass-squared $q^2 \in \left[m_\ell^2,\,(m_B - m_{D^{*}})^2\right]$, whereas for a massive invisible particle the bound shifts to $q^2 \in \left[(m_\ell + m_\chi)^2,\,(m_B - m_{D^{*}})^2\right]$, leading to characteristic changes in the decay distributions. Following the discussion,
the decay width for $B \to D^{*} \ell X_{\rm inv}$ can be written as
{\small
\begin{equation}\label{eq:total_decay_width}
	\Gamma(B \to D^{*} \ell X_{\rm inv}) = \Gamma(B \to D^{*} \ell \nu) \Bigr|_{SM} + \Gamma(B \to D^{*} \ell \chi )\Bigr|_{NP}\,.
\end{equation}
}
The detailed mathematical expressions for the decay widths $\Gamma(B \to D^{*} \ell \nu)|_{SM}$ and the respective decay distributions are available in the literature~\cite{Sakaki:2013bfa,Becirevic:2019tpx}.
The differential decay width for the process $\Gamma(B \to D^* (\to D \pi) \ell \chi)$ can be obtained by integrating the four-fold angular distribution which 
can be written as:
\begin{widetext}
 \begin{equation}\label{eq:fivefold_decaywidth}
		\frac{d}{ dm_{D\pi}^2}\left(\frac{d^4 \Gamma }{d q^2 \, d\cos \theta_{D} \,  d\cos \theta_{\ell} \, d\phi} \right) = \frac{\lambda^{1/2}(m_B^2, m_{D^*}^2, q^2)}{256 (2\pi)^6 m_{B}^3}\frac{\lambda^{1/2}(q^2, m_{\ell}^2, m_\chi^2)}{q^2} \frac{|\vec{p}_{D}|}{m_{D\pi}} \sum_{\lambda_{\ell}, \lambda_{\chi}, \lambda_{\Dst}} \left| \mathcal{M}^{\lambda_{D^*}}_{\lambda_{\ell},\lambda_{\chi}}(B\to D \pi \ell \chi ) \right|^2.
	\end{equation}
\end{widetext}
with $|\vec{p}_{D}| = \frac{\lambda^{1/2}(m_{D^*}^2, m_{D}^2, m_{\pi}^2)}{2m_{\Dst}}$ and the K{\"a}llen function is defined as $\lambda(\alpha,\beta,\gamma) = \alpha^2+\beta^2+\gamma^2-2\alpha\beta - 2\beta\gamma-2\alpha\gamma$. The corresponding SM result can then be recovered by retaining only the contribution from the operator $\mathcal{O}_{V_{1L}}^{\ell,SM}$ and taking the limit of a massless $\chi$. In the above equation $\mathcal{M}^{\lambda_{D^*}}_{\lambda_{\ell},\lambda_{\chi}}(B\to D \pi \ell \chi )$ represents the four-body helicity amplitude and the parameters $\lambda_\ell$, $\lambda_{D^*}$ and $\lambda_{\chi}$, respectively, represent the helicities of the charged lepton, $D^*$ meson and dark sector fermion $\chi$. The calculation details of this helicity amplitude can be found in detail in the ref.~\cite{Kolay:2026mgv}. In the above equation,
$\thD,\, \thl, \,$ and $\phi$ are different angles, defined in the decay plane of fig.~\ref{fig:decay_plane}.
The resulting four-fold decay distribution is then obtained as follows:
\begin{widetext}
\begin{align}
\label{eq:total_dist}
	\frac{d^4 \Gamma (B\to D^*\ell X_{\mathrm{inv}})}{dq^2 \, d \cos\thl \, d\cos\thD \, d\phi} =\frac{d^4 \Gamma (B\to D^*\ell \nu)}{dq^2 \, d \cos\theta_\ell \, d\cos\theta_D \, d\phi}
	+ \frac{d^4 \Gamma (B\to D^*\ell \chi)}{dq^2 \, d \cos\theta_\ell \, d\cos\theta_D \, d\phi}
    = \dfrac{9}{32 \pi} \times I(q^2,m_{\ell}^2, m_{\chi}^2,\theta_\ell,\thD,\phi) .
\end{align}
where,
\begin{eqnarray}
\label{eq:dGamma}
I(q^2,m_{\ell}^2, m_{\chi}^2,\theta_\ell,\thD,\phi)&=& \Big\{ I^s_{1}\, \sin^2{\thD} + I^c_{1}\, \cos^2{\thD}
+ \left( I^s_{2}\, \sin^2{\thD} + I^c_{2}\, \cos^2{\thD}\right) \cos{2\thl}
\nn \\ && 
+ \left( I_3\, \cos{2\phi} + I_9\, \sin{2\phi} \right) \sin^2{\thD} \sin^2{\thl}  
+ \left( I_4\, \cos{\phi} + I_8\, \sin{\phi} \right) \sin{2\thD} \sin{2\thl}
\nn \\ && 
+ \left( I_5\,\cos{\phi} + I_7\,\sin{\phi}  \right) \sin{2\thD} \sin{\thl}  
+ \left( I^s_{6}\, \sin^2{\thD} +  I^c_{6}\, \cos^2{\thD}\right)\cos{\thl} \Big\}.
\end{eqnarray}
\end{widetext}
This four-fold decay distribution is experimentally studied in ref.~\cite{Belle:2023xgj}, where the corresponding angular coefficients can be extracted as
{\small
\begin{equation}\label{eq:angular_coefficients}
	I_i(q^2, m_{\ell}^2, m_{X_{\mathrm{inv}}}^2)
	= I_i(q^2, m_{\ell}^2)^{B\to D^*\ell \nu }
	+ I_i(q^2, m_{\ell}^2, m_{\chi}^2)^{B\to D^*\ell \chi }.
\end{equation}
}
These angular coefficients are directly related to the hadronic helicity amplitudes. It is important to note that the kinematically allowed range of $q^2$ differs between the SM case with a massless neutrino and the scenario involving a massive dark-sector particle. 
The above-defined angular coefficients can be redefined as $\hat{I}(q^2)$ in which the systematic uncertainties are expected to cancel in the ratio. The normalized angular coefficients are defined as:
\begin{equation}
    \hat{I}_i(q^2)= \frac{	I_i(q^2, m_{\ell}^2)^{B\to D^*\ell \nu }
    	+ I_i(q^2, m_{\ell}^2, m_{\chi}^2)^{B\to D^*\ell \chi }}{\frac{d\Gamma(B\to D^*\ell X_{\mathrm{inv}})}{dq^2}},
    \label{eq:ang_normalised}
\end{equation}
In the above equation, $d\Gamma(B\to D^*\ell X_{\mathrm{inv}})/dq^2$ will be obtained from  eq.~\eqref{eq:total_dist}. 

The linear dependence on $\cos\theta_\ell$ can be understood from eq.~\eqref{eq:dGamma} via the observable forward-backward asymmetry, defined as:
\begin{equation}
\mathcal{A}^\ell_{FB}(q^2)
= \bigl(d \Gamma/ d q^2 \bigr)^{-1}(I^c_{6} + 2 I^s_{6}) \,.
\end{equation}
Integrating over $\phi$ and $\theta_\ell$ in eq.~\eqref{eq:dGamma}, the $\theta_D$ distribution becomes:
\begin{equation}
\frac{d^2 \Gamma}{d q^2 d \cos\theta_D}
= \frac{3}{4} \frac{d \Gamma}{d q^2} \left[ F_T^{D^*}(q^2)\sin^2\theta_D + 2 F_L^{D^*}(q^2)\cos^2\theta_D \right],
\end{equation}
where the polarization fractions are defined as \cite{Becirevic:2019tpx},
{\small \begin{align}
 F_T^{D^*} = \frac{2(3 I^s_{1}-I^s_{2})}{3 I^c_{1}+ 6 I^s_{1}-I^c_{2}-2 I^s_{2}}, \,\,
F_L^{D^*} = \frac{3 I^c_{1}-I^c_{2}}{3 I^c_{1}+ 6 I^s_{1}-I^c_{2}-2 I^s_{2}}.
\end{align}}

\paragraph{\underline{Analysis and Results}:}
The helicity amplitudes depend on the $B \to \Dst$ transition form factors $f_i(q^2)$. Therefore, determining the $q^2$ shapes of the hadronic matrix elements associated with the decay is essential.
To determine the $q^2$ shape of the form factors, 
we adopt Boyd-Grinstein-Lebed (BGL) \cite{Boyd:1994tt,Boyd:1997kz} parameterizations. We express the form factors $ \mathcal{F}_{i}$ as a series expansion in $z$ as: 
\begin{eqnarray}
	\mathcal{F}_i (z) = \frac{1}{P_i (z) \phi_i (z)} \sum_{j=0}^{N} a_{j}^{\mathcal{F}_i} z^j,
	\label{eq:FF-BGL}
\end{eqnarray}
where $z$ is related to the recoil angle $w$ as:
\begin{eqnarray}\label{eq:z_definition}
	z = \frac{\sqrt{w+1}-\sqrt{2}}{\sqrt{w+1}+\sqrt{2}}\,.
\end{eqnarray}
The recoil variable is related to the momentum transfer $q^2$ as $q^2 = m_B^2 + m_{\Dst}^2 - 2 m_B m_{\Dst} w$.
The Blaschke factors, $P_i (z)$, are given by
\begin{eqnarray}
	P_i(z) = \prod_p \frac{z-z_{p_{i}}}{1 - z z_{p_{i}}},
	\label{eq:Blaschke-fact}
\end{eqnarray}
which are used to eliminate the poles at $z=z_p$ where,
\begin{equation}\label{eq:Zp}
	z_p = \frac{\sqrt{(m_B + m_\Dst)^2 - m_P^2} - \sqrt{4 m_B m_\Dst}}{\sqrt{(m_B + m_\Dst)^2 - m_P^2} + \sqrt{4 m_B m_\Dst}}.
\end{equation}
Here $m_P$ denotes the pole masses below the pair production threshold. The BGL coefficients $a_{j}^{\mathcal{F}_i}$ satisfy weak unitary constraints; for the details, please see the ref. \cite{Bigi:2017jbd,Boyd:1997kz}.
Following eq.~\eqref{eq:FF-BGL}, we express the four form factors $\mathcal{F}_i = \{  f(z)$, $g(z)$, $\mathcal{F}_1 (z), \, \mathcal{F}_2 (z)\} $ relevant for $B \to D^{*} $ decay as a series expansion in $z$. The detailed mathematical expressions of $P_i (z)$s and the outer functions $\phi_i (z)$ are given in the ref.~\cite{Kolay:2026mgv}.
%
Additionally, these form factors satisfy the following QCD constraints at both zero and maximum recoil: 
\begin{subequations}
	\begin{eqnarray}
		\mathcal{F}_1 (1) &=& m_B (1-r) f(1),\\
		\mathcal{F}_2 (w_{max}) &=& \frac{1+r}{m_B^2 (1+w_{max}) (1-r)r} \mathcal{F}_1 (w_{max}). \qquad
		\label{eq:FF-kin-constr}
	\end{eqnarray}
\end{subequations}
We incorporate these constraints and use inputs from Fermilab-MILC~\cite{FermilabLattice:2021cdg}, JLQCD~\cite{Aoki:2023qpa}, HPQCD~\cite{Harrison:2023dzh}, and LCSR at $q^2=0$~\cite{Gubernari:2018wyi} to extract the BGL coefficients via a $\chi^2$ minimisation. This enables us to determine the $q^2$ dependence of the form factors and evaluate the corresponding angular observables. Details on the extracted BGL coefficients are discussed in \cite{Kolay:2026mgv}.

\begin{figure*}[t]
	\centering
	\vspace{-0.1cm}
	\subfloat[]{\includegraphics[scale=0.37]{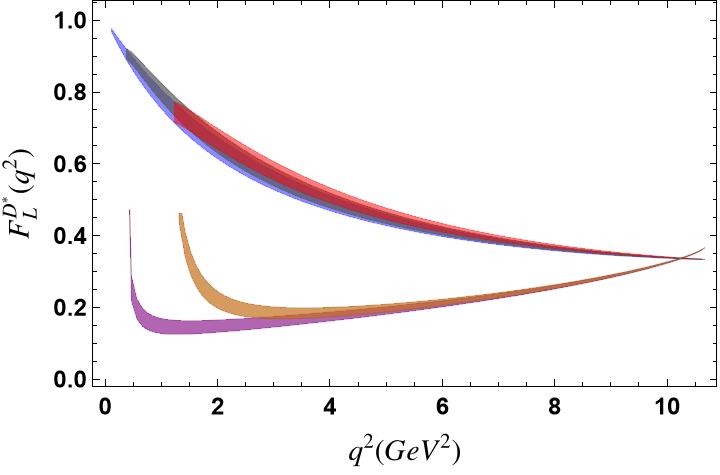}}
	\hspace{0.0005cm}
	\subfloat[]{\includegraphics[scale=0.37]{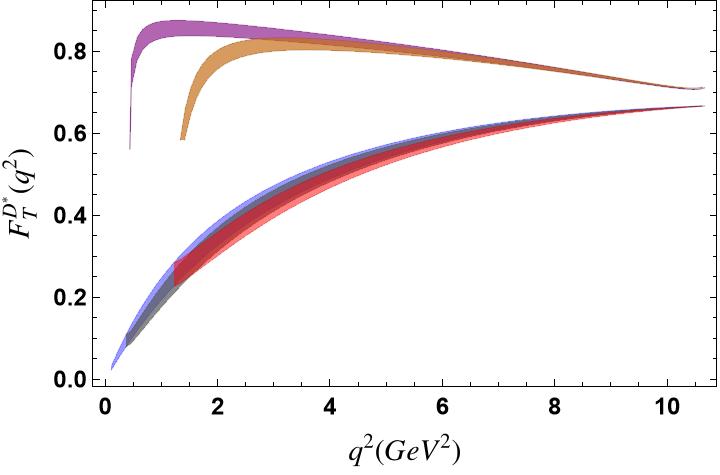}}
	\vspace{0.1cm}\hspace{0.0005cm}
	\subfloat[]{\includegraphics[scale=0.37]{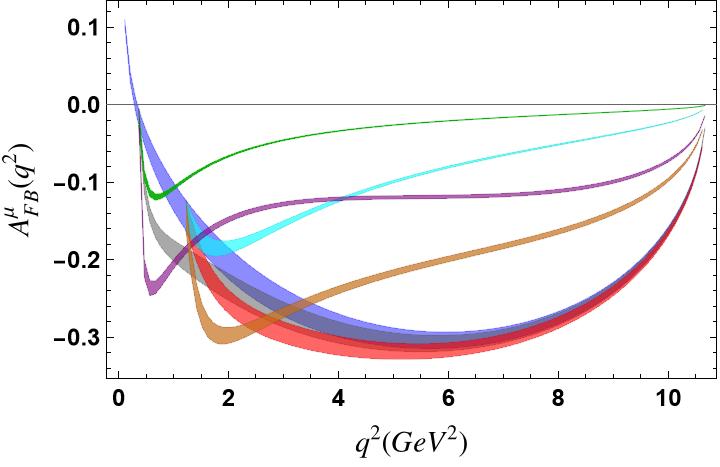}}\\
	\subfloat[]{\includegraphics[scale=0.37]{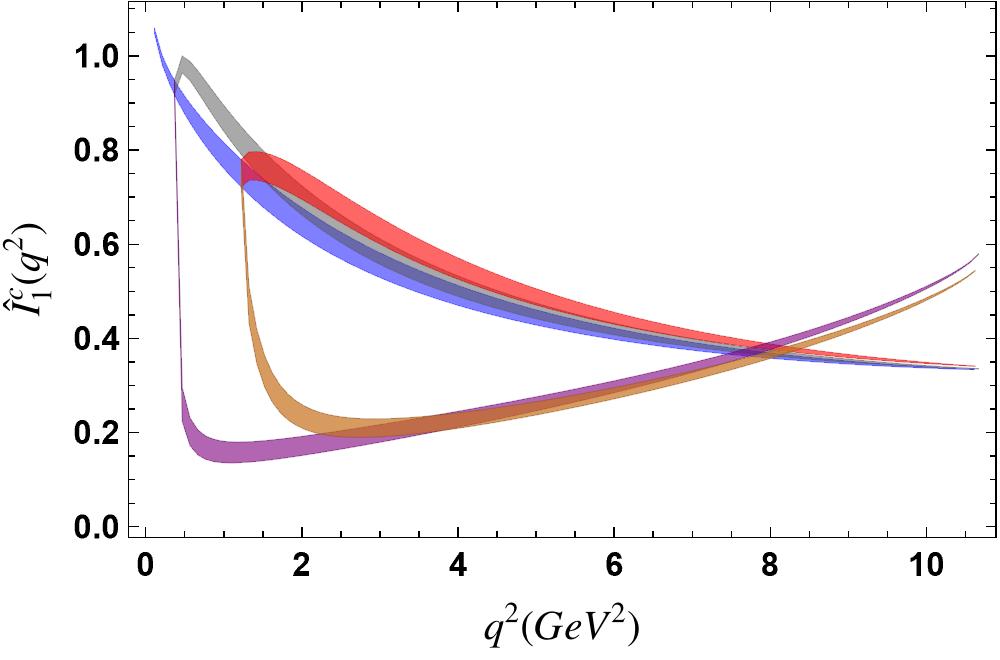}}
	\vspace{-0.1cm}\hspace{0.0005cm}
	\subfloat[]{\includegraphics[scale=0.37]{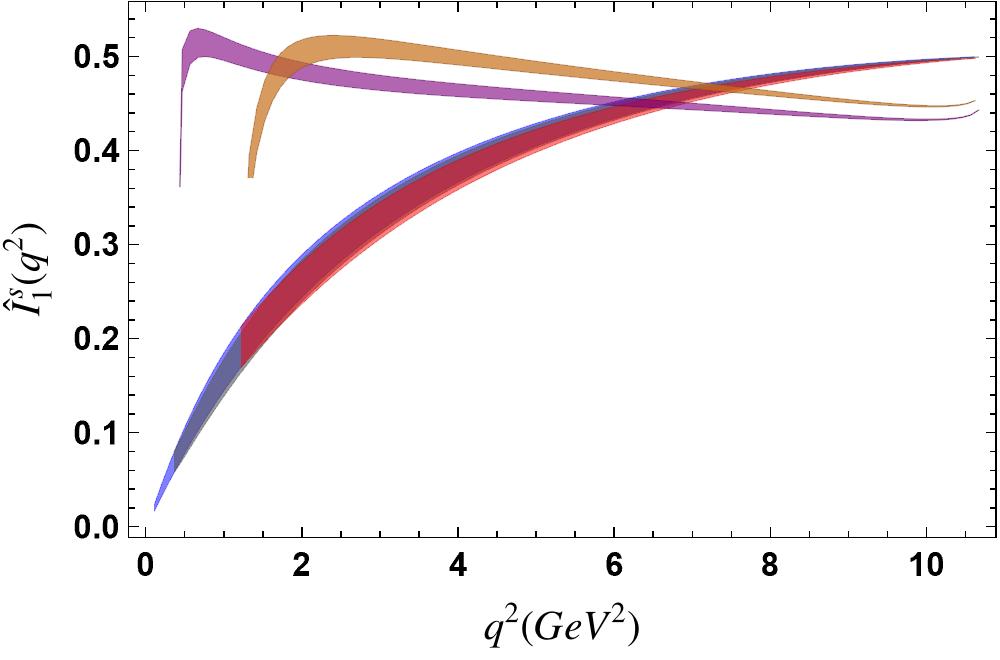}}
	\vspace{-0.1cm}\hspace{0.0005cm}
	\subfloat[]{\includegraphics[scale=0.37]{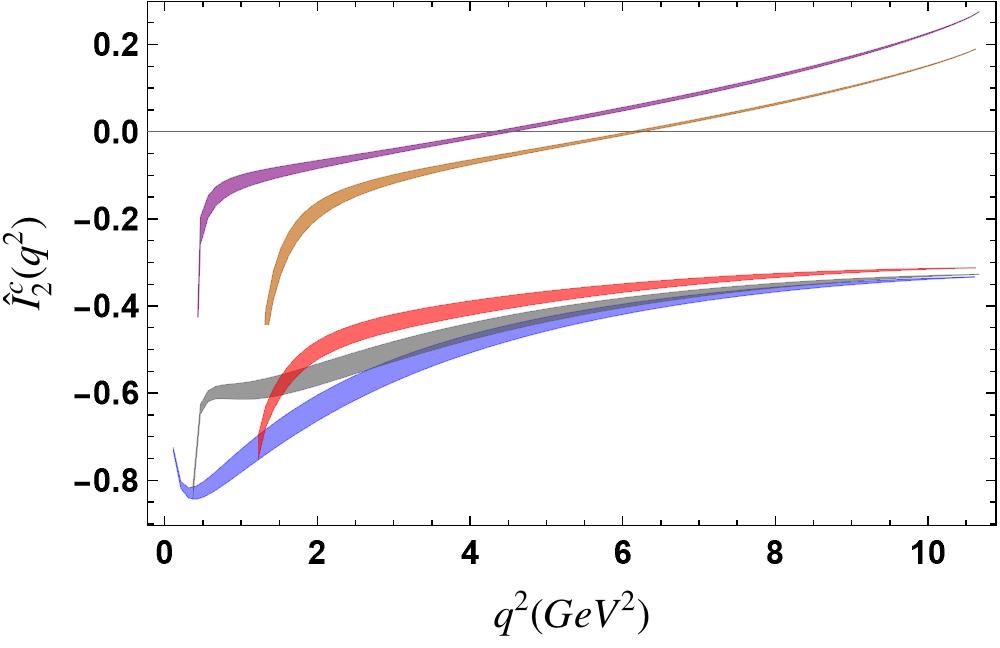}}
	\vspace{-0.1cm}\hspace{0.0005cm}
	\subfloat[]{\includegraphics[scale=0.37]{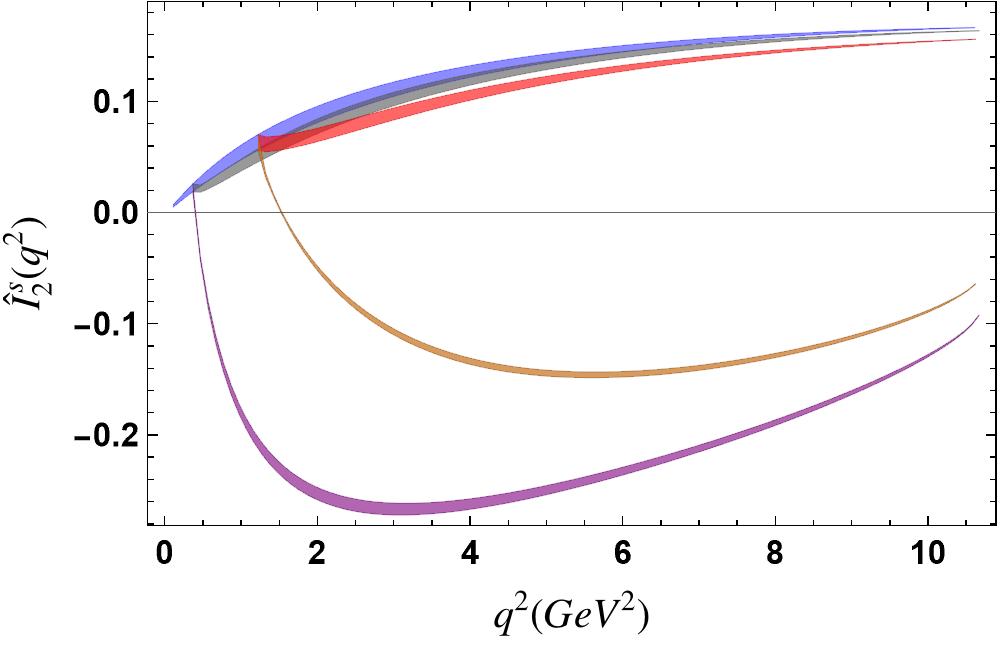}}
	\vspace{-0.1cm}\hspace{0.0005cm}
	\subfloat[]{\includegraphics[scale=0.37]{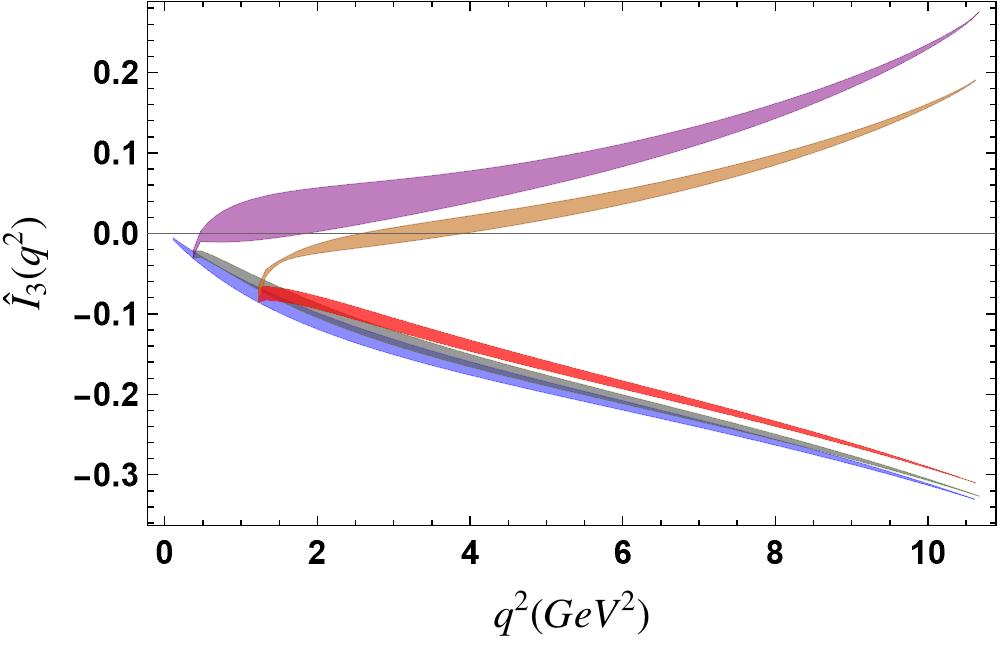}}
	\vspace{-0.1cm}\hspace{0.0005cm}
	\subfloat[]{\includegraphics[scale=0.37]{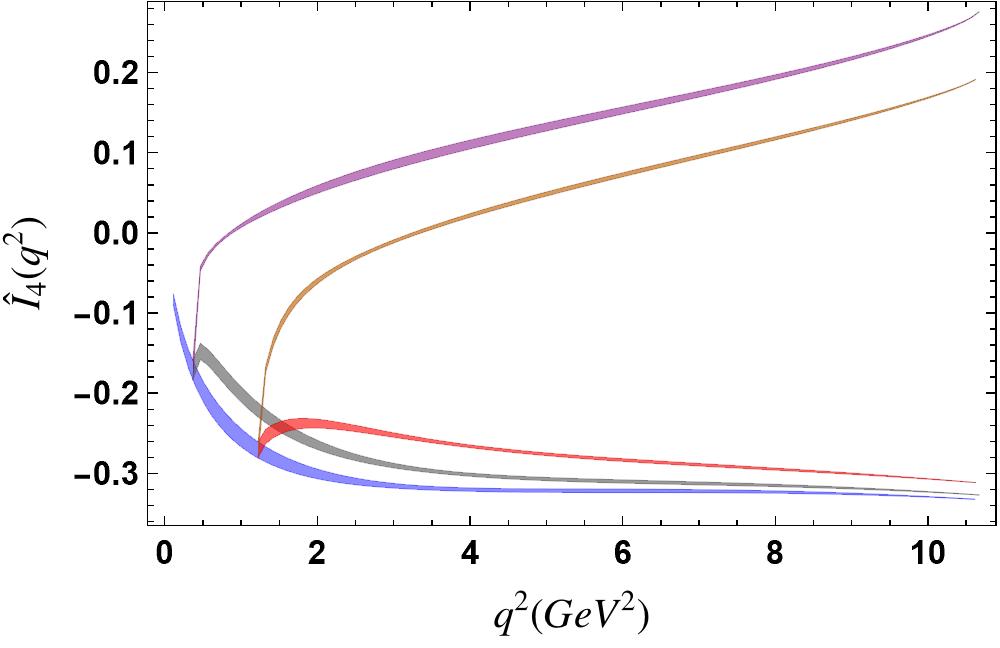}}
	\vspace{-0.1cm}\hspace{0.0005cm}
	\subfloat[]{\includegraphics[scale=0.37]{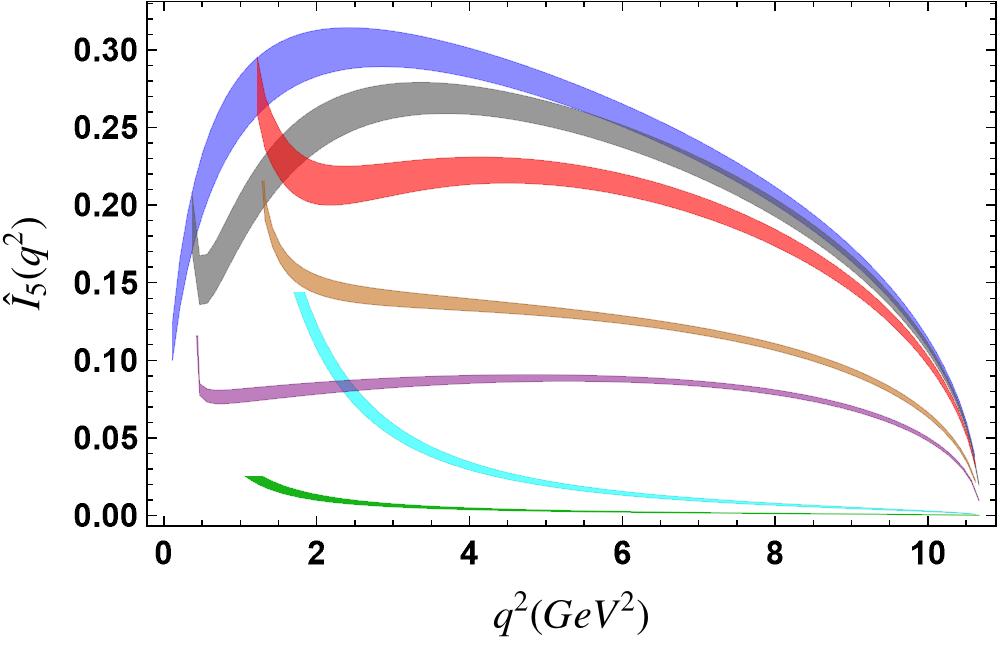}}
	\vspace{-0.1cm}\hspace{0.0005cm}
	\subfloat[]{\includegraphics[scale=0.37]{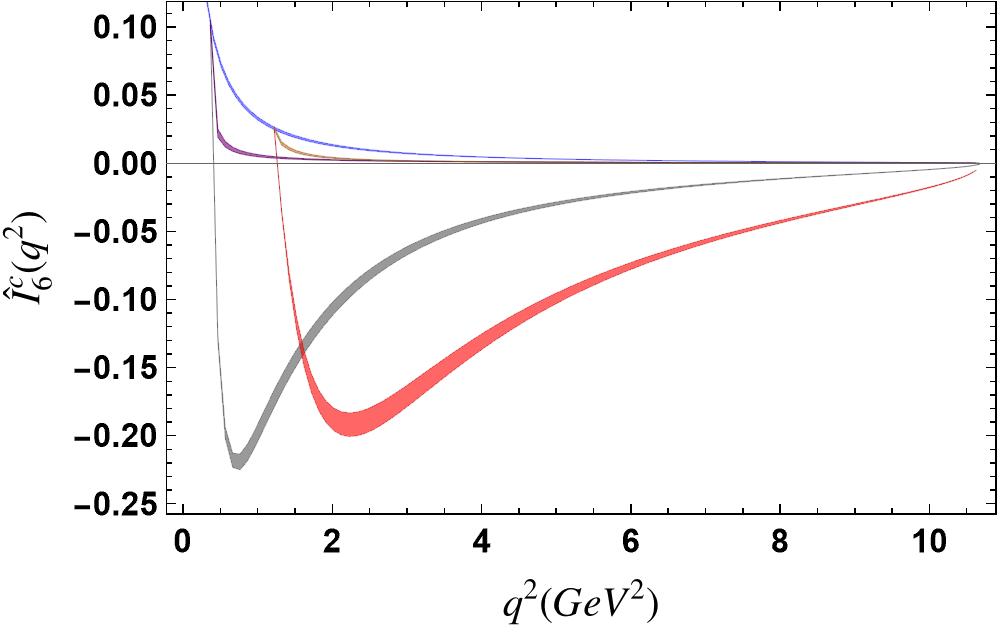}}
	\vspace{-0.1cm}\hspace{0.0005cm}
	\subfloat[]{\includegraphics[scale=0.37]{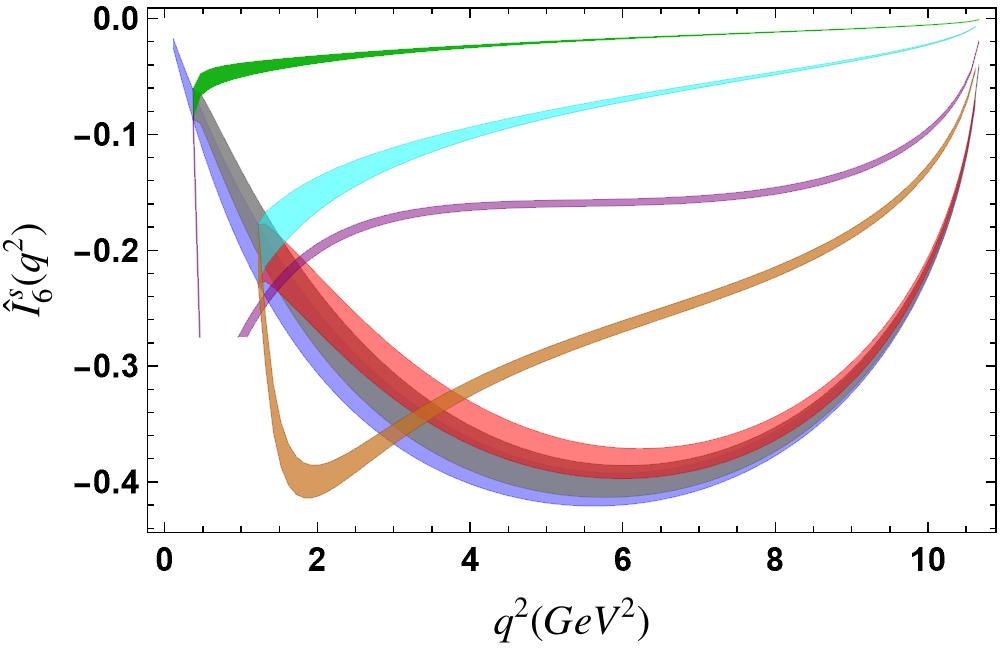}}
	\vspace{0.2cm}\hspace{0.0005cm}
	\includegraphics[scale=0.6]{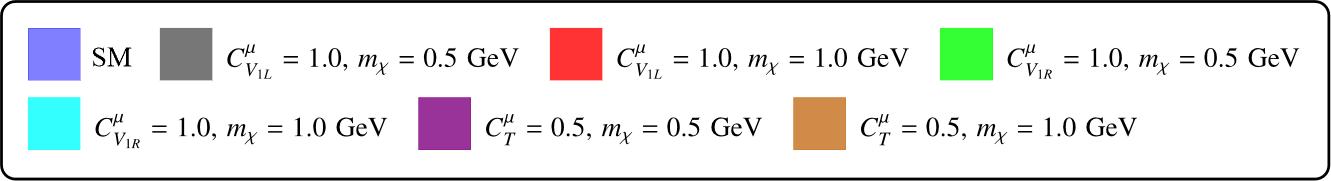}
	\caption{The $q^2$-distribution of the angular observables in the one-operator scenarios, considering both left- and right-handed currents for benchmark values of WCs and $m_{\chi}$. For non-zero mass of the invisible particle, the region $q^2 \in [m_{\ell}^2, \,  (m_{\ell} + m_{\chi})^2]$ follows purely SM distribution, hence, not shown in the plots. The colored band shows the corresponding $1\sigma$ uncertainties of the observables.}
	\label{fig:B2Dst_angular_plotsctLL}
\end{figure*}

We investigate the imprints of NP and the effects of a massive invisible particle, $\chi$, on the angular observables discussed above. Our primary goal is to identify the observables, which are genuinely sensitive to the mass of $\chi$. Not all angular observables show such sensitivity. Also, this dependence is not universal across all NP scenarios. Therefore, in this work, we focus exclusively on the subset of observables and NP frameworks where the effects of a non-zero $\chi$ mass are significant. We have a complementary objective: to isolate observables that can discriminate between left- and right-handed lepton currents. These observables are valuable for disentangling the chiral structure of the underlying interactions. They are also important for distinguishing NP scenarios involving dark-sector fermions with both left- and right-chiral couplings from frameworks such as sterile-neutrino or right-handed neutrino models, which typically have a more restricted chiral structure.

In fig.~\ref{fig:B2Dst_angular_plotsctLL}, we present the $q^2$ distributions of the angular observables that are sensitive to $m_{\chi}$ and/or can discriminate between left- and right-handed lepton currents. These distributions are shown for representative benchmark values of $m_{\chi}$, with WCs $\lesssim \mathcal{O}(1)$, consistent with current state-of-the-art constraints \cite{Kolay:2026mgv}. While the plots are largely self-explanatory, we summarise the key observations below:
\begin{itemize}[noitemsep]
    
    \item We do not observe any significant deviations in the case of scalar-scalar four-fermion operators compared to those of the SM.

    \item Almost all the angular observables listed above are highly sensitive to $\mathcal{O}^{\ell}_{T_{L,R}}$. However, none of them can distinguish between left- and right-handed tensor lepton currents. Apart from $F_{L,T}^{D^*}(q^2)$ and $\hat{I}_1^c(q^2)$, nearly all observables exhibit sensitivity to $m_{\chi}$, with the most pronounced effects seen in  $A_{FB}^{\mu}(q^2)$, $\hat{I}_2^s(q^2)$, $\hat{I}_{4,5}(q^2)$, and $\hat{I}_6^{c,s}(q^2)$ distributions.

    \item The observables $A_{FB}^{\mu}(q^2)$, $\hat{I}_5(q^2)$, and $\hat{I}_6^s(q^2)$ are highly sensitive to $\mathcal{O}^{\ell}_{{V_1}_R}$, while their sensitivity to $\mathcal{O}^{\ell}_{{V_1}_L}$ is negligible. 
    In contrast, $\hat{I}_6^c(q^2)$ is highly sensitive to both $\mathcal{O}^{\ell}_{{V_1}_{L,R}}$ and can not be distinguished.
    This complementarity provides a clear handle for distinguishing left- and right-handed lepton currents. The behaviour of $\mathcal{O}^{\ell}_{{V_2}_{L,R}}$ is analogous to that of $\mathcal{O}^{\ell}_{{V_1}_{R,L}}$, respectively.

    \item The combined measurement of these observables will be crucial in disentangling the contributions from $\mathcal{O}^{\ell}_{{V_1}_{L,R}}$ and $\mathcal{O}^{\ell}_{T_{L,R}}$, thereby providing insight into the underlying NP scenario(s).
\end{itemize}

We analyse semileptonic $B \to D^{*}\ell X_{\rm inv}$ decays as probes of light invisible fermions within a model-independent effective field theory framework, focusing on angular observables that provide information inaccessible through decay-rate measurements alone. A non-zero mass for the invisible particle induces characteristic modifications in angular distributions. We identify observables that are particularly sensitive to its mass and the chiral structure of its interactions. These observables are especially effective in distinguishing dark-sector fermions with mixed chiralities from those involving sterile or purely right-handed neutrinos. Furthermore, we demonstrate that a complete angular analysis can, in principle, unambiguously determine the nature of the underlying new physics.


\acknowledgments
LK and SS acknowledge financial support under Project Grant No. OTH/R\&D/13467 and the ANRF Grant No. ANRF/IRG/2024/000256/PS, respectively.
\bibliographystyle{ieeetr}
\bibliography{refs}
\end{document}